\documentclass[12pt,preprint]{aastex}

\shorttitle{Circular-ribbon flare} \shortauthors{Li et al.}

\begin{document}

\title{Two Episodes of Magnetic Reconnections During a Confined Circular-ribbon Flare}

\author{Ting Li\altaffilmark{1,2}, Shuhong Yang\altaffilmark{1,2}, Qingmin Zhang\altaffilmark{3}, Yijun Hou\altaffilmark{1,2} \& Jun Zhang\altaffilmark{1,2}}

\altaffiltext{1}{CAS Key Laboratory of Solar Activity, National
Astronomical Observatories, Chinese Academy of Sciences, Beijing
100101, China; liting@nao.cas.cn} \altaffiltext{2}{School of
Astronomy and Space Science, University of Chinese Academy of
Sciences, Beijing 100049, China} \altaffiltext{3}{Key Laboratory for
Dark Matter and Space Science, Purple Mountain Observatory, CAS,
Nanjing 210034, China}

\begin{abstract}

We analyze a unique event with an M1.8 confined circular-ribbon
flare on 2016 February 13, with successive formations of two
circular ribbons at the same location. The flare had two distinct
phases of UV and EUV emissions with an interval of about 270 s, of
which the second peak was energetically more important. The first
episode was accompanied by the eruption of a mini-filament and the
fast elongation motion of a thin circular ribbon (CR1) along the
counterclockwise direction at a speed of about 220 km s$^{-1}$. Two
elongated spine-related ribbons were also observed, with the inner
ribbon co-temporal with CR1 and the remote brightenings forming
$\sim$ 20 s later. In the second episode, another mini-filament
erupted and formed a blowout jet. The second circular ribbon and two
spine-related ribbons showed similar elongation motions with that
during the first episode. The extrapolated 3D coronal magnetic
fields reveal the existence of a fan-spine topology, together with a
quasi-separatrix layer (QSL) halo surrounding the fan plane and
another QSL structure outlining the inner spine. We suggest that
continuous null-point reconnection between the filament and ambient
open field occurs in each episode, leading to the sequential opening
of the filament and significant shifts of the fan plane footprint.
For the first time, we propose a compound eruption model of
circular-ribbon flares consisting of two sets of successively formed
ribbons and eruptions of multiple filaments in a fan-spine-type
magnetic configuration.

\end{abstract}


\section{Introduction}

Solar flares are believed to be caused by the reconnections of
magnetic field lines that convert magnetic energy into kinetic
energy of accelerated particles and thermal energy (Forbes et al.
2006). The morphology and dynamics of flare ribbons help to
understand the coronal magnetic reconnection process (Gorbachev et
al. 1988). Recent high-resolution observations of eruptive flares
have revealed the intrinsic 3D nature of solar flares, e.g., the
formation of twisted flux ropes (Green et al. 2011; Yang et al.
2017; Yan et al. 2017; Xue et al. 2017; Cheng et al. 2017), double
J-shaped ribbons, as well as the slipping motions of flare loops (Li
\& Zhang 2014, 2015; Dud{\'{\i}}k et al. 2014, 2016). These
complicated structures and evolution during solar flares cannot be
accommodated by the classical 2D flare model (Shibata \& Magara
2011), and thus we need to establish the 3D flare model to
understand the 3D magnetic configurations and the reconnection
process (Priest \& Forbes 2002). Until now, 3D reconnection models
for triggering solar flares have been proposed, e.g., the 3D
null-point reconnection in fan-spine topology (Lau \& Finn 1990;
Priest \& Titov 1996; Liu et al. 2011; Sun et al. 2013; Wyper et al.
2017), and the slipping reconnection model (Aulanier et al. 2012;
Janvier et al. 2013).

The topological structure of coronal magnetic fields is an important
factor in solar explosive events, and often refers to null points,
separatrix surface and separator field lines (Longcope 1996; Priest
\& Titov 1996; Wang 1997; Pontin et al. 2013). They are regions of
magnetic connectivity discontinuities and serve as preferred sites
for magnetic reconnection. The 3D null point is always embedded in a
multipolar magnetic field and associated with the fan-spine
configuration (T{\"o}r{\"o}k et al. 2009; Liu et al. 2011). The
fan-spine topology consists of a dome-shaped fan separatrix surface
dividing two different connectivity domains and a spine field
passing by the null point. The formation of 3D fan-spine
configuration is related to the flux emergence into an oblique
unipolar coronal field (Antiochos 1998; Moreno-Insertis et al.
2008). The presence of coronal fan-spine topology in coronal jets
and circular-ribbon flares is confirmed by potential and nonlinear
force-free field (NLFFF) extrapolations (Fletcher et al. 2001;
Masson et al. 2017). At the null, magnetic-breakout reconnection
process is triggered and removes the strapping field of the flux
rope beneath the fan plane (Antiochos et al. 1999; Sun et al. 2013;
Wyper et al. 2018). Then the flux rope rises up towards the null and
later reconnects with the ambient open field through null-point
reconnection, forming the coronal jets and circular flares.

Besides the reconnection at the null, magnetic reconnection in 3D
can also occur at quasi-separatrix layers (QSLs; D{\'e}moulin et al.
1996, 1997), which are the 3D generalization of separatrices and
correspond to the regions of very strong magnetic connectivity
gradients (Mandrini et al. 1997; Chandra et al. 2011). The apparent
slipping motion of field line footpoints is observed in QSL
reconnection when magnetic connectivity is continuously exchanged
between neighbouring field lines (Priest \& D{\'e}moulin 1995;
D{\'e}moulin et al. 1996). In recent years, the observational
evidences supporting the existence of slipping magnetic reconnection
have been found in several works (Dud{\'{\i}}k et al. 2014, 2016; Li
\& Zhang 2014, 2015; Zheng et al. 2016; Jing et al. 2017). Aulanier
et al. (2007) first observed the slipping motions of coronal loops
at a speed of 30-150 km s$^{-1}$. Flare loops were seen to exhibit a
quasi-periodic slipping motion with a period of 3$-$6 min (Li \&
Zhang 2015) and their slippage was in opposite directions towards
both ends of the ribbons (Dud{\'{\i}}k et al. 2016). The slipping
motion of flux rope field lines was also detected during eruptive
flares at a speed of tens of km s$^{-1}$, associated with the
elongations of flare ribbons (Li \& Zhang 2014; Li et al. 2016).

Of particular interest in the 3D magnetic reconnection regime is
circular-ribbon flare exhibiting the fan-spine topology (Masson et
al. 2009, 2017; Wang \& Liu 2012; Mandrini et al. 2014; Liu et al.
2015; Joshi et al. 2015; Zhang et al. 2016a, 2016b; Hou et al. 2016;
Hernandez-Perez et al. 2017; Hao et al. 2017; Xu et al. 2017). In
circular-ribbon flare, null-point reconnection occurs and the
reconnection-accelerated particles flow from the null along the
separatrix surface and the spine field into the lower atmosphere.
Thus circular ribbons are formed at the intersections of the fan
surface and the chromosphere (Reid et al. 2012; Wang \& Liu 2012). A
central ribbon and remote brightenings would appear inside and
outside the circular ribbon, which respectively correspond to the
footpoints of inner and outer spine. Recent observations showed that
the circular ribbon brightened sequentially and two spine-related
ribbons were elongated (Masson et al. 2009; Jiang et al. 2015).
Masson et al. (2009) suggested that a larger QSL-halo of a finite
thickness surrounds the fan and spine separatrices, and slipping or
slip-running reconnections within the QSLs cause the sequential
brightenings of flare ribbons.


The numerical simulations and NLFFF extrapolations showed the
presence of a twisted flux rope below a 3D null-point dome in the
eruptive circular-ribbon flares (Jiang et al. 2013; Yang et al.
2015; Masson et al. 2017; Li et al. 2017b). The flux rope erupted
due to the breakout-type reconnection near the null and formed a
blowout jet (Wang \& Liu 2012). Contrary to eruptive flares,
confined circular-ribbon flares do not generate any coronal mass
ejections (CMEs). Until now, the thorough dynamic evolution and
triggering mechanisms of confined circular-ribbon flares are rarely
studied. In this paper, we report a unique event. The M1.8 confined
circular-ribbon flare on 2016 February 13 had two distinct phases of
ultraviolet (UV) and extreme ultraviolet (EUV) emissions separated
by a 4-min time delay. Successive formations of two circular ribbons
at the same location and propagating brightenings of the ribbons are
first presented here. The paper is organized as follows. In Section
2, we describe the observations and data analysis. Section 3
presents the dynamic evolution of flare ribbons and overlying loops,
as well as the extrapolated magnetic topology. We summarize our
findings and discuss the interpretation of the flare in Section 4.

\section{Observations and Data Analysis}

In this event, the central flaring region containing the circular
ribbons has a small spatial scale of about
30$\arcsec$$\times$30$\arcsec$. It is fortunate that the
\emph{Interface Region Imaging Spectrograph} (\emph{IRIS}; De
Pontieu et al. 2014) detected the flare, which allows an assessment
of the detailed evolution of the flare. The \emph{IRIS} observations
provide the 1330 {\AA} slit jaw images (SJIs) with a spatial pixel
size of $\sim$0$\arcsec$.166, a field of view (FOV) of
119$\arcsec$$\times$119$\arcsec$, and a cadence of about 9.6 s. The
spine loops and remote brightenings of the circular-ribbon flare
were observed by the \emph{Solar Dynamics Observatory} (\emph{SDO};
Pesnell et al. 2012). The Atmospheric Imaging Assembly (AIA; Lemen
et al. 2012) on board the \emph{SDO} observes the Sun in 10 UV/EUV
passbands with a resolution of $\sim$0$\arcsec$.6 per pixel and a
cadence of 24$/$12 s. We mainly concentrate on the data of AIA 1600,
304, 171 and 131 {\AA} channels in this work. The Helioseismic and
Magnetic Imager (HMI; Scherrer et al. 2012) line-of-sight (LOS)
magnetograms are also used to investigate the relations of flare
ribbons with photospheric magnetic fields.

Moreover, we extrapolate the 3D coronal magnetic fields by the NLFFF
model with the optimization method (Wheatland et al. 2000;
Wiegelmann 2004). The photospheric vector magnetic field data of the
active region (AR) are observed by the HMI, with a pixel spacing of
$\sim$0$\arcsec$.5. The vector magnetogram for extrapolation is at
15:00 UT prior to the flare from the Space-weather HMI AR Patches
(SHARP; Bobra et al. 2014). Before the extrapolation, a
preprocessing procedure (Wiegelmann et al. 2006) is performed to
remove the net force and torque on the boundary. The calculation is
conducted within a box of 768$\times$512$\times$256 uniform grid
points with $\Delta$x=$\Delta$y=$\Delta$z=0$\arcsec$.5. Using the
extrapolated 3D coronal magnetic field, we calculate the squashing
factor (Q) and twist number with the code proposed by Liu et al.
(2016). The Q factor is a measurement of the deformation of the
elementary flux tube cross section, and the regions with the highest
Q values define the locations of the QSLs (Titov et al. 2002).

\section{Results}

\subsection{Overview of the Event}

On 2016 February 13, an M1.8 confined circular-ribbon flare occurred
in AR 12497 (N14, W28), which was initiated at 15:16 UT and peaked
at 15:24 UT from the GOES SXR 1$-$8 {\AA} flux data. Three flare
ribbons were distinguished from the \emph{SDO} multi-wavelength
observations, including a quasi-circular ribbon (noted CR in Figures
1(a)-(c)), an inner ribbon (IR) and remote brightenings (RB) at the
east. The comparison of 1600 {\AA} images with HMI LOS magnetograms
showed that the ribbon IR was located at the central
positive-polarity region (parasitic polarity) and CR anchored in the
surrounding negative-polarity fields (Figure 1(b)). The evolution of
photospheric magnetic fields showed that new flux emergence within
the leading negative-polarity of the AR over one day prior resulted
in the formation of this parasitic polarity. The third ribbon RB
extended northward as the flare developed, and it resided at the
trailing positive region of the AR. As seen from the
high-temperature 131 {\AA} observations (about 11 MK; O'Dwyer et al.
2010), a hot loop bundle appeared about 1 min after the formation of
RB, with its eastern end mapping to RB and the western end to the
main flaring region (Figure 1(d)). The loop bundle probably traced
the spine field lines in the fan-spine topology, and thus was name
as ``spine loops" (Sun et al. 2013).

In order to examine how the UV and EUV emissions varied with time,
we calculated the 1600 {\AA} and 304 {\AA} lightcurves of the event.
The normalised magnitudes of the lightcurves within the main flaring
region ``A" and remote brightening region ``B" are presented in
Figure 1(e). Interestingly, there are two peaks in the four
lightcurves for both areas ``A" and ``B". The 1600 {\AA} emission in
area ``A" started to increase rapidly from around 15:17:50 UT and
reached its first peak at 15:19:30 UT (black curve). Then the UV
emission decreased slightly before the second rise phase. The second
peak occurred at 15:23:40 UT, consistent with the GOES SXR 1$-$8
{\AA} flux. The time interval between the two peaks is about 4 min
and the first small peak is nearly 30 \% of the second peak. The 304
{\AA} time profile in area ``A" (green curve) has a similar
variation with the 1600 {\AA} curve. Differently, the first peak in
304 {\AA} lightcurve has a stronger emission than that in 1600 {\AA}
radiation, about 50 \% of the second peak. The emissions of the
remote brightenings RB also show two phases (red and blue curves).
The RB observed at 1600 {\AA} was formed about 20 s later than the
circular ribbon CR (red and black curves), and the first peak
occurred at 15:19:50 UT. The second phase of RB started with a 70 s
delay relative to the appearance of the second CR. The RB reached
its strongest emission at about 15:25:35 UT and then decayed
gradually. For the 304 {\AA} time profile of RB (blue curve), the
second peak has a larger emission enhancement above the background,
compared with the 1600 {\AA} profile.

\subsection{Two Episodes of the Flare Evolution}

By analyzing the \emph{IRIS} high-resolution observations in detail,
we find that two distinct episodes of energy release are involved in
the impulsive phase of the flare. Before the onset of the flare, two
filament structures (F1 and F2) existed along the eastern and
southern portions of the quasi-circular polarity inversion line
(PIL) (Figures 2(a) and 2(l)). The lengths of F1 and F2 were about
30$\arcsec$, the typical length of mini-filaments (Hermans \& Martin
1986; Hong et al. 2017; Shen et al. 2017). The northern ends of F1
and F2 were located at the central positive-polarity magnetic fields
and their southern ends anchored in the neighboring
negative-polarity fields. From about 15:18:10 UT, some faint
brightenings of the first circular ribbon (CR1) started to appear at
the northeast part (see Animation 1330-flare). Then the brightenings
of CR1 were enhanced and exhibited a slow elongation motion towards
the south (Figure 2(b)). Meanwhile, the brightenings of the first
inner ribbon (IR1) evolved to the south, similar to the dynamic
evolution of CR1. About 1 min after the appearance of CR1, the
filament F1 was brightened and began to rise up with a velocity of
about 110 km s$^{-1}$ (Figure 2(c)). The eastern end of F1 exhibited
an apparent slipping motion and the western end was located at the
vicinity of the IR1 (Figures 2(d)-(f)). These successively
brightened structures of F1 delineated a curved surface (Figure
2(f)), implying the existence of a QSL surrounding the filament F1
(Li \& Zhang 2014; Li et al. 2017c). Associated with the fast
expansion of F1, the elongation motion of CR1 became faster in the
counterclockwise direction (red arrows in Figures 2(c)-(e)). As the
CR1 traveled south, the brightenings of CR1 clearly jumped to the
west (see Figures 2(b)-(f) and Animation 1330-flare). This suggested
that the filament F1 had been reconnected with open field via
null-point reconnection and the brightenings of CR1 followed the
sequential opening of F1.

From about 15:21:03 UT, the north part of the filament F2 was
associated with evident brightenings (BF2 in Figure 2(g)), implying
the onset of the second episode of magnetic reconnections. About 20
s later, the brightenings of the second circular ribbon (CR2)
started to appear at the periphery of BF2 (Figure 2(g)). The
filament F2 underwent a slipping eruption with the south end
propagating towards the west at a speed of about 40 km s$^{-1}$
(Figures 2(g)-(h)). The eruption of F2 was also accompanied by a
counterclockwise rotating motion. Simultaneously, the circular
ribbon CR2 and the second inner ribbon (IR2) were both brightened
sequentially in the counterclockwise direction with respective
speeds of 60 and 40 km s$^{-1}$ (Figures 2(g)-(k)). The dynamic
evolution of CR2 is consistent with that of CR1, while the emission
intensity and the width of CR2 are much larger than those of CR1. In
the late phase, the erupting F2 formed a curtain-like, blowout jet
after 15:23:59 UT (Figures 2(i)-(k)). The jet seemed to consist of
thread-like fine structures, which probably traced out the fan loops
in the fan-spine topology (Figure 2(k)). The fan loops also
exhibited an apparent slipping motion towards the west and their
footpoints corresponded to the evolving bright dots within CR2.
Associated with the westward motion of the CR2, its brightenings
along the south side shifted significantly northward (compared with
the former faint brightenings at the south in Figure 2(h)). This
indicated that the filament F2 was reconnected with open field via
null-point reconnection and thus the fan plane separatrix footprint
followed the path of F2 (Wyper et al. 2017).

To present how the 1330 {\AA} emission varied in time, we calculated
the lightcurves within the main flaring region ``C" and its
subregion ``D" (Figure 2(j)) and plotted the normalised magnitudes
of the lightcurves in Figure 3(a). From about 15:18:20 UT, the
emission within area ``C" increased rapidly, corresponding to the
appearance of CR1 and the initiation of the first episode of
magnetic reconnections. The first phase lasted about 130 s, with the
peak at about 15:19:20 UT (first blue arrow). The second peak
occurred at about 15:23:50 UT (second blue arrow), which was
separated by a 270 s time delay. The second phase had stronger
emissions than the first phase, with the second peak flux nearly a
factor of 2.5 larger. The curves ``C" and ``D" were very similar and
both had two distinct phases, implying that the inner ribbons were
co-temporal with the circular ribbons. In order to analyze the
elongation motions of the circular ribbons, we obtain two stack
plots (Figures 3(b)-(c)) respectively along slice ``S2" in 1330
{\AA} images (red curve in Figure 2(i)) and slice ``S1" in 304 {\AA}
images (blue curve in Figure 1(c)). As seen from the 1330 {\AA}
stack plot, the initial elongation motion of CR1 was relatively slow
in the first tens of seconds. Then the ribbon CR1 extended rapidly
toward the south with a larger velocity of about 220 km s$^{-1}$.
About 1 min after the disappearance of CR1, CR2 was generated and
subsequently propagated in the counterclockwise direction at a
smaller velocity of 60 km s$^{-1}$. The 304 {\AA} stack plot showed
a similar result that CR1 propagated at a velocity nearly a factor
of three faster compared to CR2.

\subsection{Dynamics of the Overlying Loops}

In the 171 {\AA} and 131 {\AA} images, many long and high coronal
loops were observed above the flare (Figure 4 and Animation
EUV-loops). The HMI magnetograms and 171 {\AA} images showed that
the eastern footpoints of the loop bundle anchored in the
positive-polarity fields at the northeast of remote brightenings RB
and their western footpoints in the peripheral negative-polarity
fields of the circular ribbons (Figures 4(a) and (e)). Two loop
structures ``A" and ``B" were traced to investigate their dynamic
evolution (Figures 4(a)-(c)). Associated with the ascent of the
blowout jet, loop ``A" was pushed upward and its projected height
increased by about 9 Mm in about 8 min (Figures 4(a)-(b)). While
loop ``B" showed the oscillation pattern as seen from the stack plot
along slice ``L1" (Figures 4(c)-(d)). The oscillation had an average
period of 4 min, in agreement with the typical periods of several
minutes for the kink oscillations of coronal loops (Liu \& Ofman
2014; Zimovets \& Nakariakov 2015; Li et al. 2017a). The rising
speed of the jet was estimated to be 140$-$160 km s$^{-1}$ and the
maximum height traced was up to 40 Mm (Figure 4(d)). Due to strong
constraints of overlying loops, the erupting materials from the
blowout jet finally fell back to the solar surface with a velocity
of about 50 km s$^{-1}$. By examining the data from the Large Angle
and Spectrometric Coronagraph (LASCO) Experiment, we found that the
circular-ribbon flare did not produce any CMEs and evolved into a
confined event. In 131 {\AA} images, the upward expansion of bushy
high-temperature coronal loops was detected above the jet front
(Figures 4(f)-(g)). The displacement of these expanding loops
reached about 38 Mm and the speed was about 100 km s$^{-1}$, smaller
than the rising speed of ejecting materials (Figure 4(h)).
Eventually, the high-temperature loops gradually cooled down and can
not be discerned in 131 {\AA} images.

\subsection{Fan-spine Magnetic Topology of Flaring Region}

The extrapolation results show the existence of a fan-spine topology
prior to the flare (Figure 5). The fan structure divides the coronal
volume in two connectivity domains, the inner and the outer
connectivity domain. The field lines in inner domain are connected
with the central parasitic positive-polarity region (yellow lines in
Figure 5(a)), and the field lines in outer domain link with remote
positive-polarity region (blue lines). As seen from the distribution
of the Q factor (Figure 5(b)), a dome-shaped QSL structure (referred
to as a ``QSL-halo" in Masson et al. 2009) outlining the fan plane
and another QSL structure (IQ) outlining the inner spine inside the
dome are present. Under the fan dome, two flux ropes (FR1 and FR2)
are present respectively along the eastern and southern parts of the
quasi-circular PIL (green and red lines in Figure 5(c)). FR1
consists of moderately twisted field lines with the average twist
number of 1.1 (Berger \& Prior 2006; Liu et al. 2016). Compared to
Figure 2, FR1 bears a good resemblance to the observed filament F1.
The south flux rope FR2 is weakly twisted with the average twist
number of about 0.6, which probably corresponds to the south part of
the filament F2. In the top view of the Q map (Figure 5(d)), the
intersection of the QSL-halo with the lower boundary forms a
quasi-circular morphology that approximately overlaps the location
of circular ribbons CR1 and CR2 (Figures 2 and 5). Besides the two
QSL structures (QSL-halo and IQ) around the fan-spine topology, two
flux-rope-related QSLs (FQ1 and FQ2) exist inside the QSL-halo,
respectively binding FR1 and FR2. The combination of the
extrapolated results and the observations shows that the curved
surface delineated by the eruption of F1 may correspond to the FQ1
(Figures 2(f) and 5(d)) and the slipping motion of F2 during the
second episode is probably along the FQ2 (Figures 2(g)-(h) and
5(d)).

\section{Summary and Discussion}

We have examined a special M1.8 circular-ribbon flare on 2016
February 13. The flare accompanied by the eruptions of multiple
filaments and exhibited two distinct episodes of magnetic
reconnections. The UV and EUV emissions of the flaring region showed
two peaks with an interval of about 270 s, of which the second peak
was energetically more important. The first episode was associated
with the eruption of a mini-filament F1 and the fast elongation
motion of a thin circular ribbon CR1. The eastern end of the
erupting F1 exhibited an apparent slipping motion towards the south,
with the rising speed reaching about 110 km s$^{-1}$. Accompanied by
the fast expansion of F1, the ribbon CR1 brightened sequentially in
the counterclockwise direction at a rapid speed of about 220 km
s$^{-1}$. The two spine-related ribbons (IR1 and RB) were
simultaneously elongated. In the second episode, another
mini-filament F2 erupted and the second circular ribbon CR2 was
formed at the same location of CR1. The broad ribbon CR2 and inner
ribbon IR2 also showed counterclockwise elongation motions with
respective speeds of 60 and 40 km s$^{-1}$. The erupting F2 finally
developed into a blowout jet and the thread-like fine structures of
the jet traced out the fan loops. The jet caused the upward
expansion of the high-temperature coronal loops above the flare at a
speed of about 100 km s$^{-1}$. Due to the strong constraints of
overlying coronal loops, the flare became a confined event and did
not generate any CMEs.

The extrapolated 3D coronal magnetic fields reveal the existence of
a fan-spine topology. A dome-shaped QSL-halo outlining the fan
surface and another QSL structure around the inner spine are
present. The intersection of the QSL-halo with the lower boundary is
approximately co-spatial with circular ribbons CR1 and CR2. Two flux
ropes and two flux-rope-related QSLs exist under the fan dome, which
roughly correspond to the two filaments F1 and F2 and two respective
QSLs binding them. In each episode, the sequential brightenings of
the circular ribbon and two spine-related ribbons indicate the
shifts of the fan plane footprint and the footpoints of both spines.
A relevant simulation study was carried out by Wyper et al. (2016),
who showed that during a confined eruption the inner and outer
spines shift rapidly across the photosphere (see their Figure 11 and
the related movie), concurrent with a shifting of the fan plane
footprint. In that study, the field beneath the fan plane was
sheared, rather than including a filament, however those authors
have recently updated their model to include a filament (Wyper et
al. 2017, 2018). A similar shift in spine and fan footpoints occurs
as the filament erupts and is reconnected on to open field. Our
observations are similar to their simulation studies, and show that
the erupting filaments probably lead to significant shifts of the
fan plane footprint via the null-point reconnection.

By comparing our results with the numerical studies of Wyper et al.
(2017, 2018), we propose a compound eruption model of
circular-ribbon flares consisting of two sets of successively formed
ribbons and eruptions of multiple filaments in a fan-spine-type
magnetic configuration. A simple schematic picture is shown in
Figure 6. First, the null-point reconnection between the outermost
flux of the filament F1 (blue curves) and the ambient open field
occurs (panel (a)), which transfers partial flux of F1 to the closed
field under the far side of the dome and to the open field exterior
to the near side of the dome (dashed cyan lines in panel (a)).
Meanwhile, the circular ribbon CR1 and inner ribbon IR1 (orange
curves) are generated due to the flow of reconnection-accelerated
particles from the null along the fan plane and the spine field into
the lower atmosphere. The null-point reconnection is likely of the
breakout nature (Antiochos et al. 1999; Wyper et al. 2017, 2018).
Then the F1 undergoes the slipping magnetic reconnection along the
QSL binding the F1, which induces the far flux of F1 reaching the
null point (panel (b)). The continuous null-point reconnection
causes the sequential opening of F1, and thus leads to the
significant shifts of the brightenings of CR1 and IR1 (orange
regions in panel (b)). In the first episode, partial flux is
transferred from under the separatrix dome via the null-point
reconnection, which effectively alleviates the constraint of the
nearby filament F2 (green curves). Later on, the stability of F2 is
disrupted and the second episode of the flare is initiated. The
continuous null-point reconnection between F2 and open field is at
work and generates the sequential brightenings of CR2 and IR2
(panels (c)-(d)), similar to the evolution of the first episode.
Meanwhile, the blowout jet is formed as filament material below the
separatrix is ejected along field lines outside the separatrix.

Besides the null-point reconnection, we suggest that slipping
magnetic reconnection along the extended QSL-halo is also involved
in the flare. The elongation speeds of the two circular ribbons (60
and 220 km s$^{-1}$) are sub-Alfv\'{e}nic, satisfying the slipping
reconnection regime according to Aulanier et al. (2006). Pontin et
al. (2016) analyzed the distribution of Q in the presence of
magnetic nulls and demonstrated that the extended QSL-halo was a
generic feature for a non-radially-symmetric null. Slipping
reconnection along the QSL-halo is a natural element of 3D
null-point reconnection and occurs any time 3D null reconnection
occurs (Masson et al. 2009, Reid et al. 2012). We suggest that the
null-point and slipping magnetic reconnections are both at work in
the two episodes of the flare, with the second episode being more
violent and releasing more magnetic energy. Slipping reconnection
leads to a continuous change of connectivity across QSL, while
null-point reconnection leads to a jump of connectivity and thus to
a flux transfer through separatrices.

\acknowledgments {We thank Sophie Masson and Hua-Ning Wang for
useful discussions. \emph{SDO} is a mission of NASA's Living With a
Star Program. \emph{IRIS} is a NASA small explorer mission developed
and operated by LMSAL with mission operations executed at NASA's
Ames Research center and major contributions to downlink
communications funded by the Norwegian Space Center (NSC, Norway)
through an ESA PRODEX contract. This work is supported by the
National Natural Science Foundations of China (11773039, 11533008,
11790304, 11673035, 11773079, 11673034 and 11790300), Key Programs
of the Chinese Academy of Sciences (QYZDJ-SSW-SLH050), and the Youth
Innovation Promotion Association of CAS (2017078, 2017368 and
2014043).}

{}
\clearpage

\begin{figure}
\centering
\includegraphics
[bb=32 166 541 608,clip,angle=0,scale=0.8]{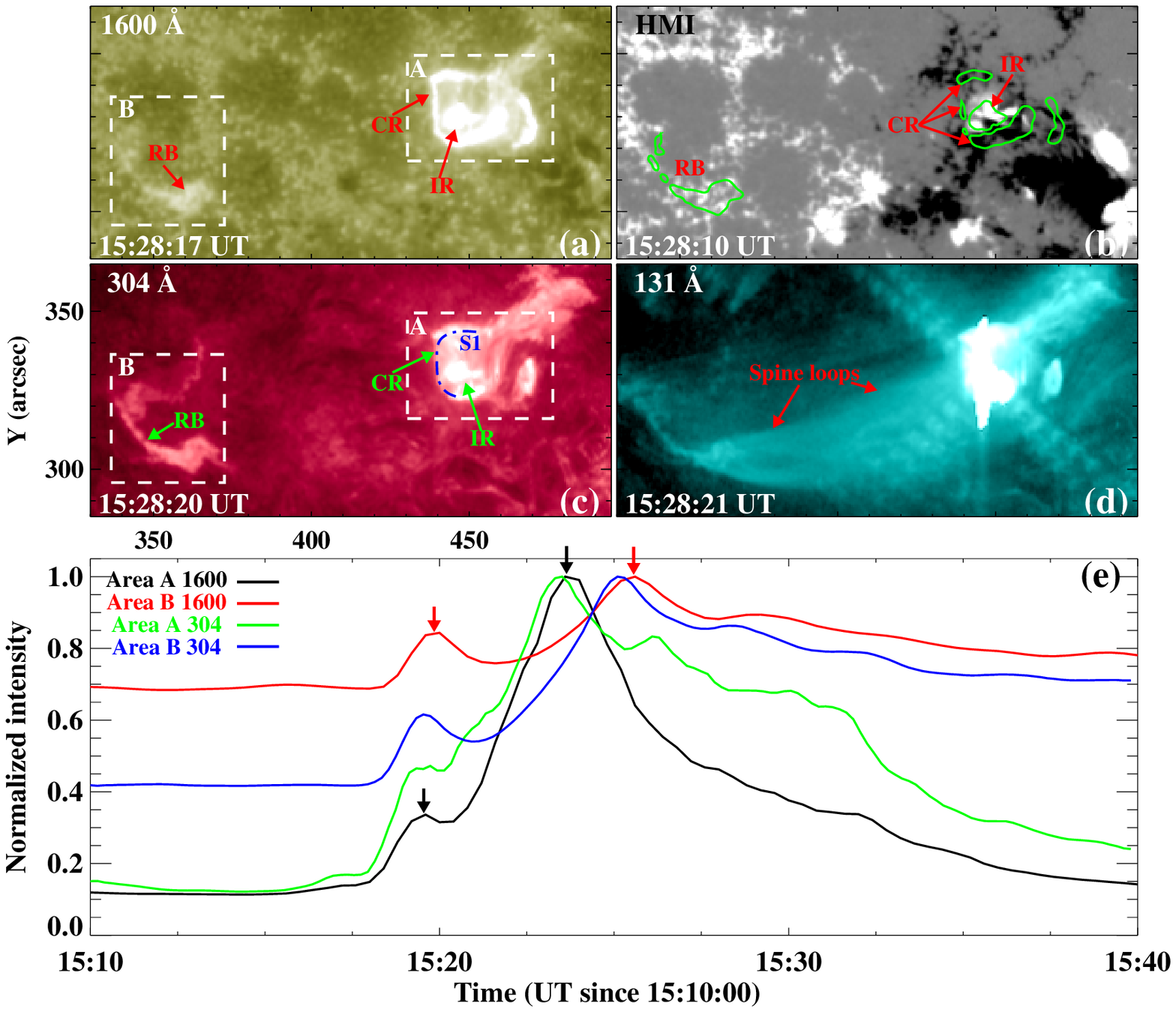}
\caption{Overview of the M1.8 circular-ribbon flare on 2016 February
13 in AR 12497 including the multi-wavelength images from
\emph{SDO}/AIA, \emph{SDO}/HMI LOS magnetogram, and lightcurves of
the flare. CR, IR and RB respectively denote the circular ribbon,
inner ribbon and remote brightenings. The two areas A and B in
panels (a) and (c) enclose the flare core region and the remote
brightening region. The green curves in panel (b) are the brightness
contours of the three ribbons in the AIA 1600 {\AA} image. The blue
dash-dotted curve ``S1" in panel (c) shows the cut position used to
obtain the time-distance plot shown in Figure 3(c). The lightcurves
in panel (e) are normalised to 1, and black and red arrows point to
two peaks of the 1600 {\AA} lightcurves within areas A and B.
\label{fig1}}
\end{figure}
\clearpage

\begin{figure}
\centering
\includegraphics
[bb=12 115 553 701,clip,angle=0,scale=0.8]{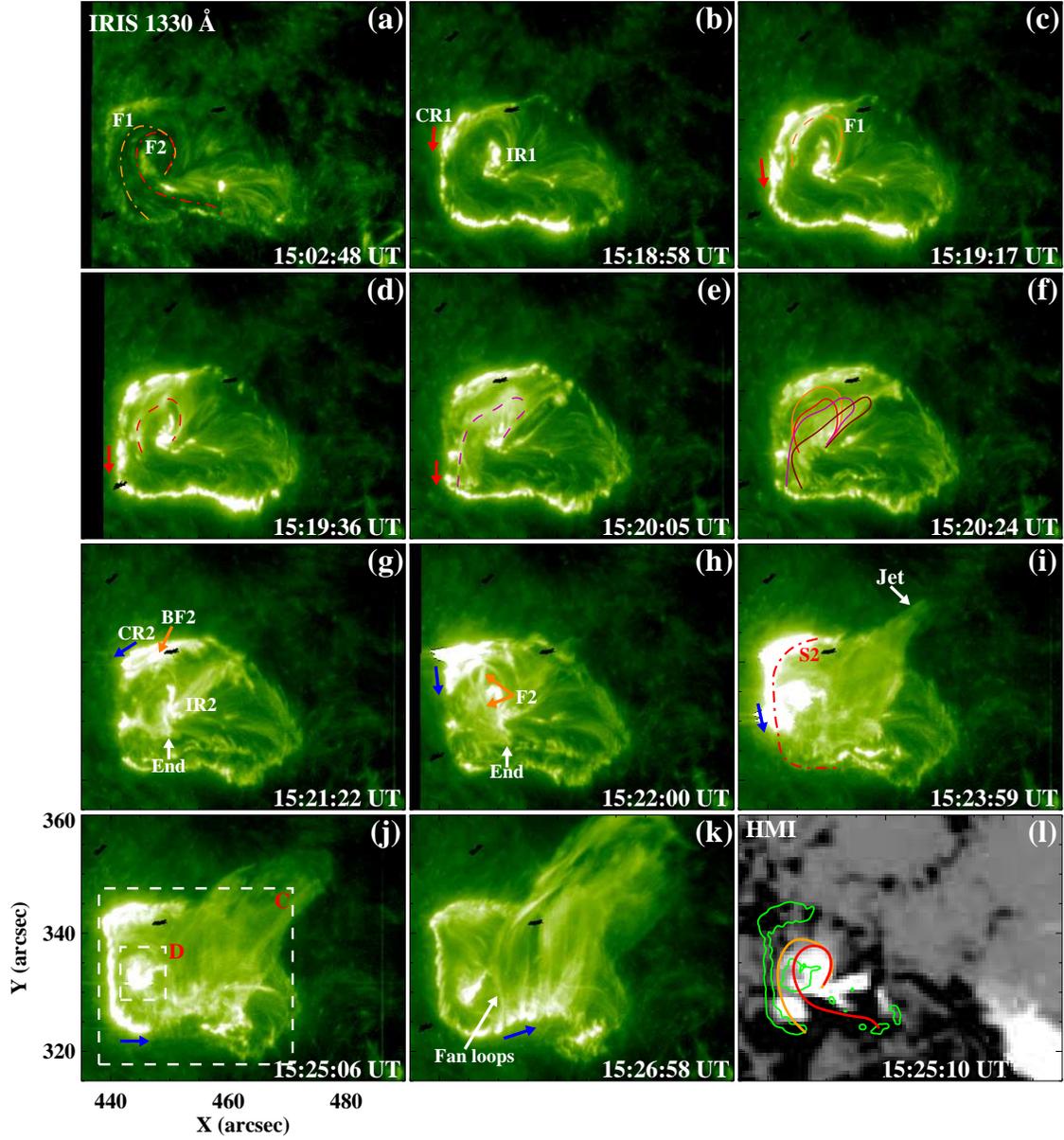} \caption{Time
sequences of \emph{IRIS} 1330 {\AA} images showing two episodes of
the evolution of circular ribbons (see Animation 1330-flare). Two
dash-dotted curves in panel (a) outline two filament structures F1
and F2 prior to the flare, and their duplications are also plotted
in panel (l). CR1 and IR1 in panel (b) denote the first circular and
inner ribbons, and red arrows in panels (b)-(e) represent the
elongation motion of CR1. The dashed and solid curves in panels
(c)-(f) denote the successive appearance of the erupting F1. CR2 and
IR2 in panel (g) are the second circular and inner ribbons, and blue
arrows in panels (g)-(j) indicate the elongation motion of CR2. BF2
in panel (g) denotes the initial brightening of F2. The red
dash-dotted curve ``S2" in panel (i) shows the cut position used to
obtain the time-distance plot shown in Figure 3(b). Areas C and D in
panel (j) mark the FOVs within which the lightcurves in Figure 3(a)
are obtained. The green curves in panel (l) are the brightness
contours of CR2 and IR2 at 15:25:06 UT. \label{fig2}}
\end{figure}
\clearpage

\begin{figure}
\centering
\includegraphics
[bb=91 155 476 661,clip,angle=0,scale=0.8]{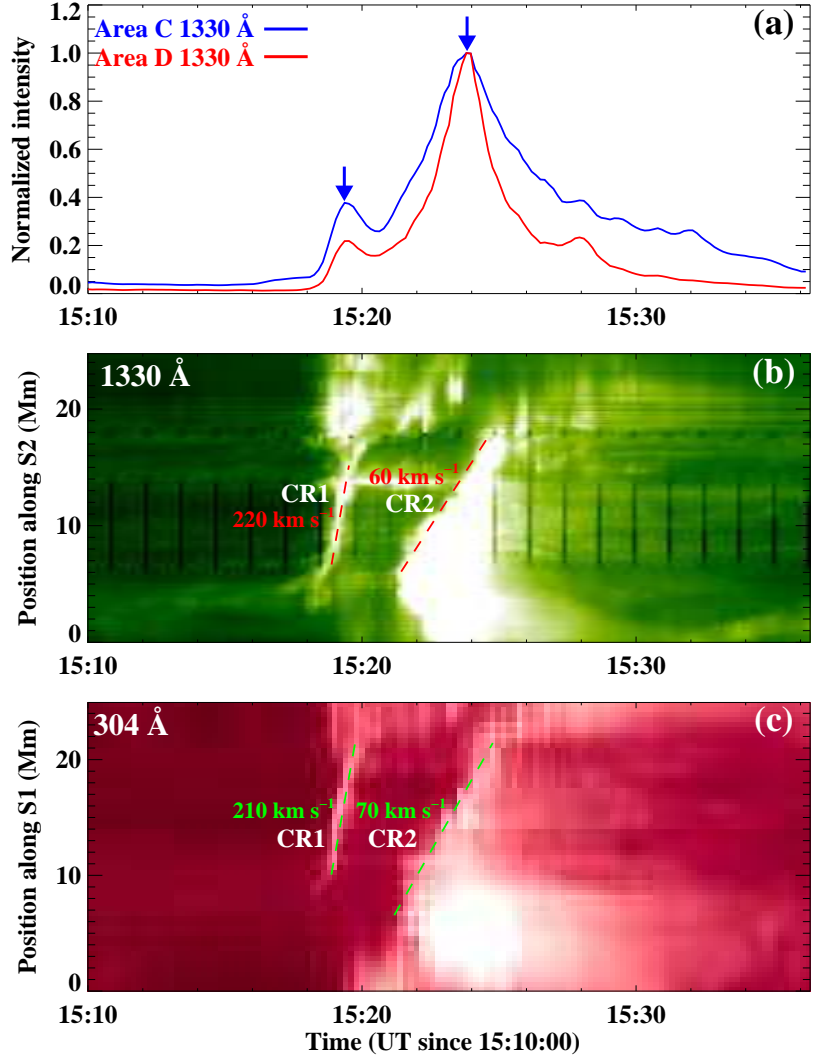} \caption{Panel
(a): normalised lightcurves of \emph{IRIS} 1330 {\AA} within areas C
and D (Figure 2(j)). Blue arrows point to two peaks of the emission
within area C. Panel (b): time-distance plot along slice ``S2" (red
dash-dotted curve in Figure 2(i)) in the 1330 {\AA} passband. Panel
(c): time-distance plot along slice ``S1" (blue dash-dotted curve in
Figure 1(c)) in the 304 {\AA} passband. \label{fig3}}
\end{figure}
\clearpage

\begin{figure}
\centering
\includegraphics
[bb=37 139 536 664,clip,angle=0,scale=0.85]{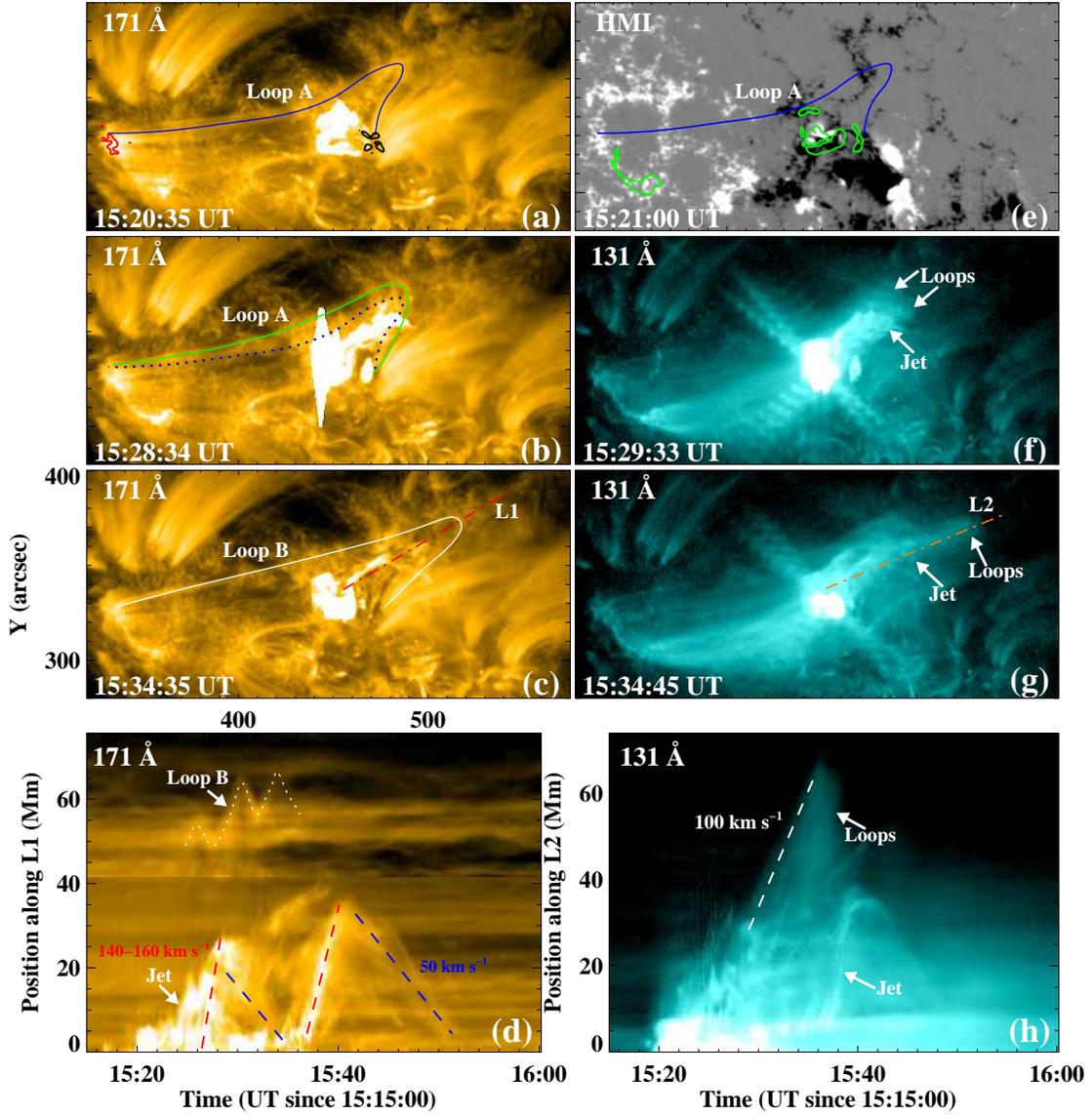}
\caption{\emph{SDO}/AIA 171 {\AA} and 131 {\AA} observations showing
the expansions and oscillations of coronal loops overlying the
flaring region (see Animation EUV-loops). The blue and green curves
in panels (a)-(b) and (e) outline the locations of loop ``A" at
different times. White solid and dotted curves in panels (c)-(d)
denote the loop ``B" and its oscillation behavior. Red and dark
contours in panel (a) are the magnetic fields at $\pm$400 G levels
at the ends of loop ``A". Panel (d) shows the time-distance plot
along slice ``L1" (red dash-dotted line in panel (c)) in the 171
{\AA} passband. The red and blue dashed lines in panel (d)
respectively indicate the ejecting and falling materials from the
blowout jet. The green contours in panel (e) are the brightness of
flare ribbons in the 1600 {\AA} channel. Panel (h) shows the
time-distance plot along slice ``L2" (brown dash-dotted line in
panel (g)) in the 131 {\AA} passband. The white dashed line in panel
(h) denotes the expansion motion of the overlying loops.
\label{fig4}}
\end{figure}
\clearpage

\begin{figure}
\centering
\includegraphics
[bb=43 262 524 539,clip,angle=0,scale=0.85]{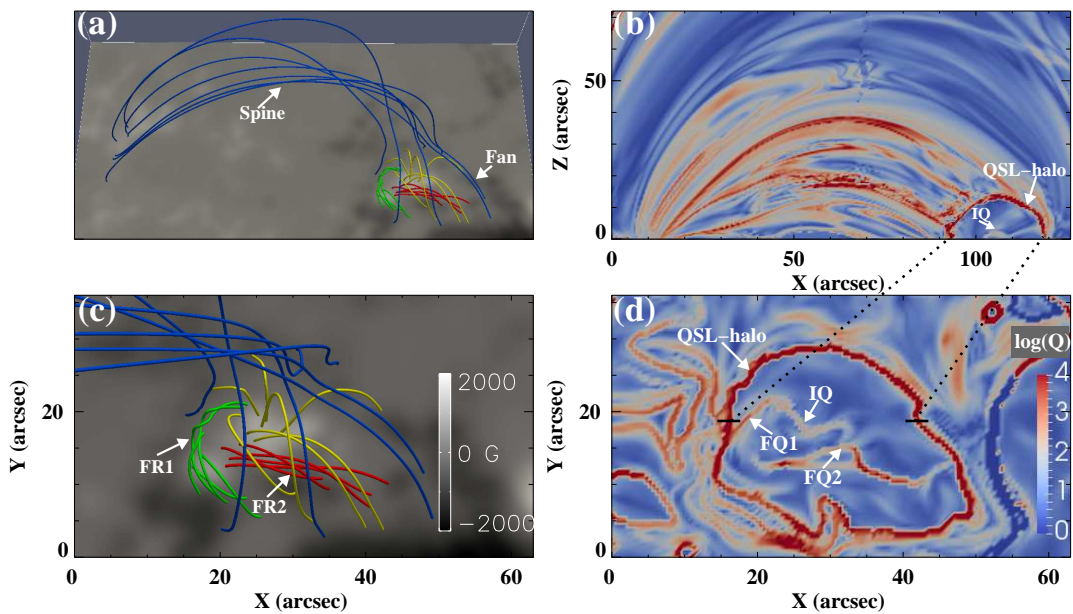} \caption{Panel
(a): side view of the extrapolated field lines at 15:00 UT showing
the fan-spine topology. The background shows the photospheric
vertical magnetic field. Panel (b): distribution of the squashing
factor Q in the x-z plane obtained from the NLFFF field, which
depicts the QSL-halo around the separatrix (fan) and the IQ
surrounding the inner spine field lines. Panel (c): top view of the
zoomed image of panel (a). FR1 and FR2 are two flux ropes underlying
the fan dome. Panel (d): logarithm Q on the bottom boundary. FQ1 and
FQ2 denote two flux-rope-related QSLs within the QSL-halo.
\label{fig5}}
\end{figure}
\clearpage

\begin{figure}
\centering
\includegraphics
[bb=1 259 585 667,clip,angle=0,scale=0.75]{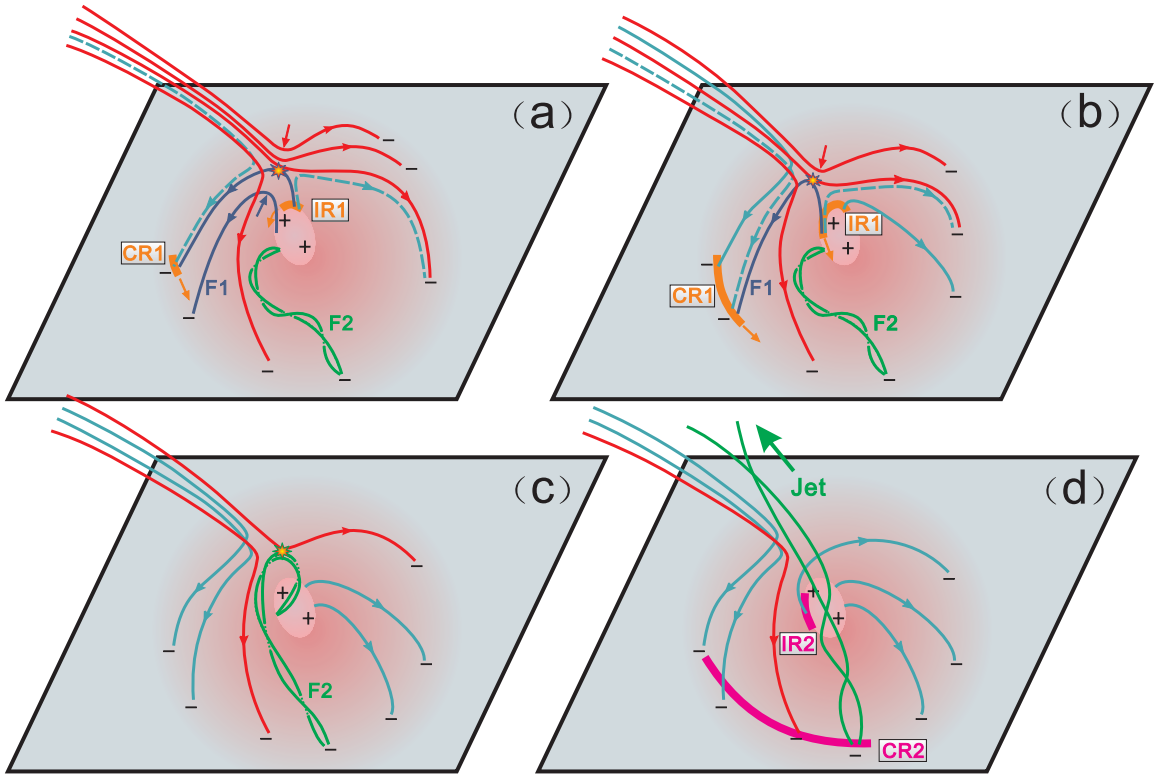}
\caption{Schematic of the compound eruption process. Red lines show
the open field lines of the fan dome. Blue and green lines
respectively represent the two filaments F1 and F2 under the
separatrix. At the null, the filament and ambient open field lines
undergo breakout-type reconnection. Dashed cyan lines denote the
newly formed open fields exterior to the near side of the dome and
the closed fields under the far side of the dome. The circular
ribbon CR1 and inner ribbon IR1 in the first episode are shown by
orange in panels (a)-(b). The orange arrows represent the sequential
brightenings of CR1 and IR1. The magenta lines in panel (d) denote
the circular ribbon CR2 and inner ribbon IR2 in the second episode.
\label{fig6}}
\end{figure}
\clearpage

\end{document}